\documentclass[12pt]{iopart}

\usepackage{graphicx}
\begin{document}

\title[Preliminary study for anapole moment measurements in rubidium and francium]{Preliminary studies for anapole moment measurements in rubidium and francium.}

\author{D. Sheng and L. A. Orozco}

\address{Joint Quantum Institute, Department of Physics, University of Maryland and National Institute of Standards and Technology, College
Park, MD 20742-4111, USA}
\ead{lorozco@umd.edu}

\author{E. Gomez}

\address{Instituto de F\'isica, Universidad Aut\'onoma de San Luis Potos\'i, San Luis Potos\'i, SLP 78290, M\'exico }

\begin{abstract}
Preparations for the anapole measurement in Fr indicate the possibility of performing a similar measurement in a chain of Rb. The sensitivity analysis based on a single nucleon model shows the potential for placing strong limits on the nucleon weak interaction parameters. There are values of the magnetic fields at much lower values than found before that are insensitive to first order changes in the field. The anapole moment effect in Rb corresponds to an equivalent electric field that is eighty times smaller than Fr, but the stability of the isotopes and the current performance of the dipole trap in the apparatus, presented here, are encouraging for pursuing the measurment.

\end{abstract}

\maketitle

\section{Introduction}

The constraints obtained from Atomic Parity Non-Conservation (PNC) on the weak interaction and its manifestation both at low energy and in hadronic environments are unique \cite{behr09}.  The information it provides is complementary to that obtained with high energy experiments. The last twenty years have seen steady progress in the experimental advances \cite{vetter95,wood97,guena05b,tsigutkin09} together with atomic theoretical calculations \cite{dzuba89,blundell92,safronova00,wansbeek08,sahoo06} having a precision better than 1\% \cite{porsev09,dzuba06,ginges04}. 

As we prepare for a new generation of PNC experiments with radioactive isotopes \cite{gomez06,gomez07}, we continue to study the measurement strategy and find advances both in the understanding of the parameters and on the specific experimental approaches that we are taking to achieve the ultimate goal. This paper presents progress on both fronts. We explore the possibility of measurements in Rb and Fr in our current apparatus and show the calculated anapole moments using current single particle nuclear models. We find the possible constraints in the nuclear weak interaction parameters that such measurements can bring and present the current performance of our atomic trap.

The general approach for the PNC experiments under consideration is the interference between an allowed transiton and the weak interacting PNC transition \cite{bouchiat74,bouchiat74b}. These experiments are going to take place in atomic traps and will require access to accelerators that can deliver the different isotopes.

\section{The anapole moment in atoms}

We start reviewing the basics of the anapole moment following very closely the work we have done in planning the experiment in Fr \cite{gomez07}. Parity nonconservation in atoms appears through two types of weak interaction: Nuclear
spin independent and nuclear spin dependent \cite{bouchiat97}. Nuclear spin dependent
PNC occurs in three ways \cite{zeldovich59,flambaum84,ginges04}: An electron interacts weakly
with a single valence nucleon (nucleon axial-vector current $A_nV_e$), the nuclear
chiral current created by weak interactions between nucleons (anapole moment), and the
combined action of the hyperfine interaction and the spin-independent $Z^0$ exchange
interaction from nucleon vector currents $(V_nA_e)$. \cite{haxton01,johnson03,dmitriev04}.

Assuming an infinitely heavy nucleon without radiative corrections, the
Hamiltonian is \cite{khriplovich91}:

\begin{equation}
H=\frac{G}{\sqrt{2}}(\kappa_{1i}\gamma_5-\kappa_{nsd,i} \mbox{\boldmath$ \sigma_n\cdot
\alpha$})\delta({\bf r}), \label{singlehamiltonian}
\end{equation}
where $G =10^{-5}$ m$_p^{-2}$ is the Fermi constant, m$_p$ is the proton mass,
$\gamma_5$ and ${\bf \alpha}$ are Dirac matrices, ${\bf \sigma_n}$ are Pauli matrices,
and $\kappa_{1i}$ and $\kappa_{nsd,i}$ (nuclear spin dependent) with $i=p,n$ for a
proton or a neutron are constants of the interaction. At tree level
$\kappa_{nsd,i}=\kappa_{2i}$, and in the standard model these constants are given by

\begin{eqnarray}
& & \kappa_{1p}=\frac{1}{2}(1-4\sin^2\theta_W), \kappa_{1n}=-\frac{1}{2}, \nonumber \\
& & \kappa_{2p}=-\kappa_{2n}=\kappa_2= -\frac{1}{2}(1-4\sin^2\theta_W)\eta,
\label{kappa}
\end{eqnarray}
with $\sin^2\theta_W \sim 0.23$ the Weinberg angle and $\eta=1.25$. $\kappa_{1i}$
($\kappa_{2i}$) represents the coupling between nucleon and electron currents when an
electron (nucleon) is the axial vector.

The first term of Eq. \ref{singlehamiltonian} gives a contribution that is independent of the nuclear 
spin and proportional to the weak charge $(Q_{W})$ in the approximation of the shell model with a single valence nucleon of unpaired spin.
For the standard model values, the weak charge is almost equal to minus the number of neutrons  ${
N}$ that we take to be proportional to the number of protons $Z$. The second term is nuclear spin dependent and due to the pairing of nucleons its
contribution has a weaker dependence on $Z$  \cite{flambaum97}:
\begin{equation}
H_{PNC}^{nsd}=\frac{G}{\sqrt{2}} \frac{K \mbox{\boldmath$I \cdot \alpha$}}{I(I+1)}
\kappa_{nsd} \delta(r), \label{hpnc}
\end{equation}
where 
\begin{equation}
K=(I+1/2)(-1)^{I+1/2-l},
\label{K}
\end{equation}
with $l$ the nucleon orbital angular momentum, $I$ is
the nuclear spin. The terms proportional to the anomalous magnetic moment of the
nucleons and the electrons are neglected.

The interaction constant is then:
\begin{equation}
\kappa_{nsd}=\kappa_a- \frac{K-1/2}{K}\kappa_2+ \frac{I+1}{K}\kappa_{Q_W},
\label{kappatotal}
\end{equation}
where $\kappa_2 \sim -0.05$ from Eq. \ref{kappa} is the tree level
approximation, and we have two corrections, the effective constant
of the anapole moment $\kappa_a$, and $\kappa_{Q_W}$ that is
generated by the nuclear spin independent part of the electron
nucleon interaction together with the hyperfine interaction.

The contributions to the interaction constant can be estimated by \cite{flambaum84,flambaum97,flambaum85}:
\begin{eqnarray}
& & \kappa_a= \frac{9}{10} g \frac{\alpha \mu}{m_p \tilde{r_0}}{\cal A}^{2/3},
\nonumber
\\ & & \kappa_{Q_W}=-\frac{1}{3} Q_W \frac{\alpha \mu_N}{m_p \tilde{r_0}{\cal A}}{\cal A}^{2/3},
\label{kappaa}
\end{eqnarray}
where $\alpha$ is the fine structure constant, $\mu$ and $\mu_N$ are the magnetic
moment of the external nucleon and of the nucleus respectively in nuclear magnetons,
$\tilde{r_0}=1.2~$fm is the nucleon radius, ${\cal A}={Z}+{ N}$, and $g$ gives
the strength of the weak nucleon-nucleon potential with $g_p \sim 4$ for protons and
$0.2<g_n<1$ for neutrons \cite{khriplovich91}. 
For a heavy nucleus like francium, the anapole moment contribution is the dominant one ($\kappa_{a,p}/\kappa_{Q_W}=14$ and $\kappa_{a,p}/\kappa_{2,n}=9$). Rubidium is heavy enough that the anapole moment contribution still dominates ($\kappa_{a,p}/\kappa_{Q_W}=20$ and $\kappa_{a,p}/\kappa_{2,n}=5$).

Vetter {\it et al.} \cite{vetter95} set an upper limit on the anapole moment of Thallium and  Wood {\it et
al.}\cite{wood97,wood99} measured with an uncertainty of $15\%$ the anapole moment of $^{133}$Cs
by extracting the dependence of atomic PNC on the hyperfine levels involved. The results form atomic PNC and other measurements in nuclear physics have similar uncertainty, but do not completely agree with each other \cite{haxton01}. It is desirable to have other atomic PNC measurements to resolve the discrepancy. In particular, in the method proposed the anapole moment dominates over the spin independent PNC contribution. The projected signal to noise is 60 times higher than that of the Cs measurement \cite{gomez07}. Measurements in ions have also been proposed \cite{fortson93,geetha98}.
Following Khriplovich {\cite{khriplovich91} the anapole moment is:
\begin{equation}
{\bf a}=-\pi \int d^3r r^2 {\bf J(r)}, \label{definea}
\end{equation}
with ${\bf J}$ the electromagnetic current density. 
 
Flambaum, Khriplovich and
Sushkov \cite{flambaum84} by including weak interactions between
nucleons in their calculation of the nuclear current density,
estimate the anapole moment from Eq. \ref{definea} of a single
valence nucleon to be
\begin{equation}
{\bf a} =\frac{1}{e}\frac{G}{\sqrt{2}}\frac{K\mbox{\boldmath$j$}}{j(j+1)}\kappa_{a} =
C^{an} \mbox{\boldmath$j$}, \label{anapolemoment}
\end{equation}
where $j$ is the nucleon angular momentum. 
The calculation is based on the shell model for the nucleus, under
the assumption of homogeneous nuclear density and a core with zero
angular momentum leaving the valence nucleon carrying all the
angular momentum. Dimitrev and Telitsin \cite{dmitriev97,dmitriev00} have looked into many body effects in anapole moments and find strong compensations among many-body contributions, still there is about a factor of two difference with the single particle result (Eq. \ref{anapolemoment}) \cite{dmitriev04}. 

We estimate with Eqs. \ref{kappaa} and \ref{anapolemoment} 
the anapole moments of five francium isotopes on the neutron deficient side with approximately one minute lifetimes and five rubidium isotopes that lie on both sides of the stability region. Eq. \ref{anapolemoment} is still a good choice to give qualitative and quantative guideline of the anapole  moment measurement. We have studied the Fr isotopes extensively in Ref.~\cite{grossman99}. In
even-neutron isotopes, the unpaired valence proton generates the
anapole moment, whereas in the odd-neutron isotopes both the
unpaired valence proton and neutron participate.  In the latter case, one must
add vectorially the contributions from both the proton and the
neutron to obtain the anapole moment as follows. 

\begin{equation}
{\bf a} =\frac{C^{an}_p \mbox{\boldmath$j$}_p\cdot \mbox{\boldmath$I$}+C^{an}_n
\mbox{\boldmath$j$}_n\cdot \mbox{\boldmath$I$}}{\mbox{\boldmath$I$}^2}
\mbox{\boldmath$I$},\label{addition}
\end{equation}
with $C^{an}_i \mbox{\boldmath$j$}_i$ the anapole moment for a
single valence nucleon $i$ in a shell model description as given
by Eq. \ref{anapolemoment}, with  the appropriate values of $j_p$ and $j_n$ depending on the isotope and using $g_p$=4 and $g_n$=1. Then we can use as an operational definition for the anapole moment constant the following equation:
\begin{equation}
{\bf a} =\frac{1}{e}\frac{G}{\sqrt{2}}\frac{(I+1/2)}{I(I+1)}\kappa_{a}{\mbox{\boldmath$I$}},
\label{operationalkappa}
\end{equation}
This way of defining the anapole moment absorbs the angular momentum constant $K$ from Eq. \ref{K} in $\kappa_a$. 

Figure \ref{isotopes}a shows the effective constant for the anapole moment for five different isotopes of francium (triangles) and rubidium (open squares). Fr has an
unpaired $\pi h_{9/2}$ proton for all the isotopes considered; the odd neutron in  $^{208,210}$Fr is an $\nu f_{5/2}$ orbit, while 
in  $^{212}$Fr the extra neutron is on a $\nu p_{1/2}$ orbital. 
There is a clear even-odd neutron number alternation in Fr due to the pairing of neutrons. For Rb, the alternation is no longer evident due to changes in the orbitals for the valence nucleons.
In particular the value of $\kappa_a$ has a different sign for the two stable isotopes of rubidium ($^{85}$Rb and $^{87}$Rb).
The nucleon orbitals used for rubidium are $\pi f_{5/2}$ for isotopes 84 and 85, $\pi p_{3/2}$ for 86-88, $\nu g_{9/2}$ for 84 and 86 and $\nu f_{5/2}$ for 88 \cite{brownpersonal}. 
The two neutron holes in $^{85}$Rb deform the nucleus very slightly and change the order of the proton orbitals from $\pi p_{3/2}$ in $^{87}$Rb to $\pi f_{5/2}$ in $^{85}$Rb. The result is that the spin and orbital contributions to the angular momentum are anti-aligned in $^{85}$Rb and they are aligned in $^{87}$Rb. The alignment is responsible for the sign change in $\kappa_a$.
The authors of Ref.~\cite{angstmann05} use Eq.~\ref{kappa} to calculate the anapole moment constant and find no sign change
between $^{87}$Rb and $^{85}$Rb. We consider even and odd isotopes with the
vector sum of Eq.~\ref{addition}. The sign change that we get comes from 
the contribution of $K$ (Eq.~\ref{K}) in our operational definition of the anapole moment Eq.~\ref{operationalkappa}. The quantity measured experimentally, the amplitude of the $E1$ PNC
transition, also contains the sign change.

Figure \ref{isotopes}b presents a sensitivity analysis of the effective anapole constant for the Rb isotopes to the change in $g_p$=4 and $g_n$=1 by fifty percent up and down around the values used in Fig.\ref{isotopes}a. The range of values still preserves the basic structure of the plot, and should allow a study of the $g_n/g_p$ ratio. There still remains the question of the sensitivity to the 
configuration used for the particular nucleus. 
The calculation of the anapole constant uses the orbital expected to be the dominant one. The actual orbital may be a different one or even a superposition of different orbitals. Using a proton orbital $\pi$p$_{3/2}$ for $^{86}$Rb changes $\kappa_a$ from 0.45 to -0.13, while using a proton orbital $pi$d$_{5/2}$ for $^{88}$Rb gives a smaller change from -0.06 to 0.01.
Rb is a  tractable nucleus as it is around the neutron magic number of 50. This is not the case for Cs where the nuclear structure calculations are more complicated.

\begin{figure}
\leavevmode \centering
\includegraphics[width=4in]{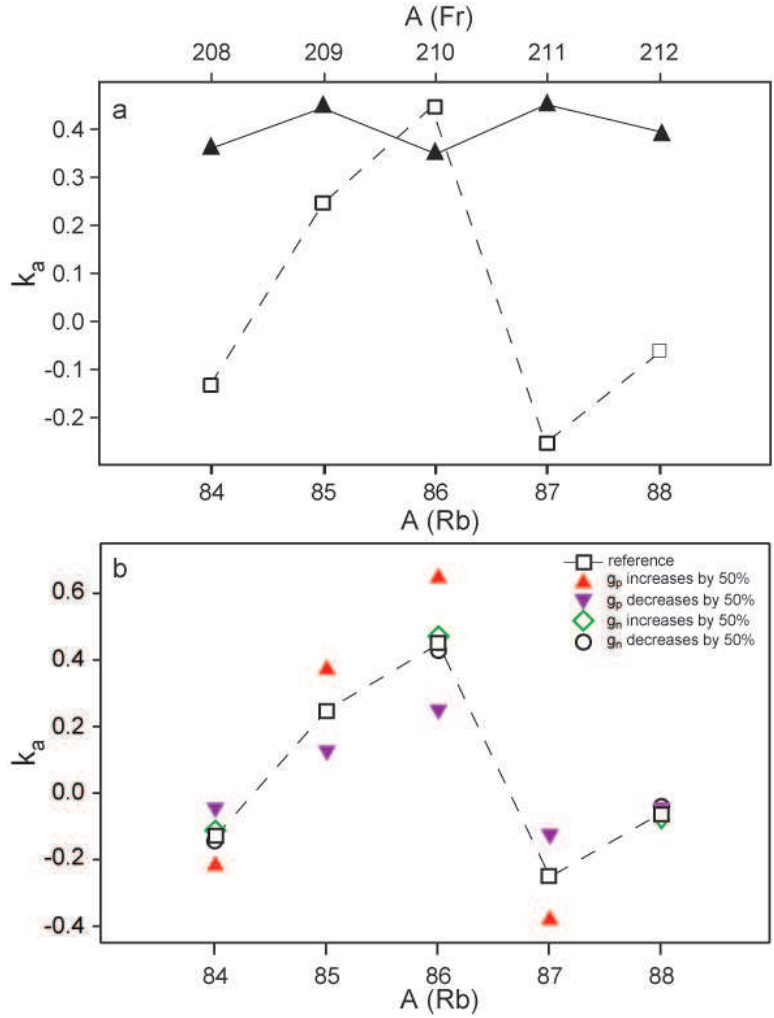}
\caption{(Color online.) Anapole moment effective constant for different isotopes. a) 
francium (triangles) and rubidium (open squares) with $g_p=4,~g_n=1$. b) Sensitivity analysis for the Anapole moment effective constant for different Rb isotopes. The limits come from varying $g_p$ and $g_n$ by fifty percent from the values in a). The lines are only to guide the eye.  \label{isotopes}}
\end{figure}

\section{Constraints to weak meson-nucleon interaction constants from anapole measurements}

The anapole moment constant ($\kappa_a$) depends on the strength of the weak nucleon-nucleus potential, characterized by $g_p$ for a proton and $g_n$ for a neutron. Equation 18 of Ref. \cite{flambaum97} gives a relation between the weak nucleon-nucleus constants ($g_p$ and $g_n$) that appear in the expression for the anapole moment  (Eq.~\ref{kappaa}) and the meson-nucleon parity nonconserving interaction constants formulated by Desplanques, Donoghue, and  Holstein (DDH) \cite{desplanques80}. Evaluating the relations with  the DDH ``best" values for the weak meson-nucleon constants we arrive at the following expressions
\begin{eqnarray}
& & Y=3.61(-X + 1.77 g_p +0.26), \label{weakconstantsa} \\
& & Y=2.5(X + 2.65 g_n -0.29), \label{weakconstantsb}
\end{eqnarray}
with $X=(f_{\pi}-0.12h_{\rho}^1-0.18h_{\omega}^1) \times 10^7$ and $Y=-(h_{\rho}^0 + 0.7 h_{\omega}^0) \times 10^7$ combinations of weak meson-nucleon constants. Figure \ref{constraints} shows the expected constraints on the weak meson-nucleon constants from an anapole moment measurement using Eqs. \ref{weakconstantsa} and \ref{weakconstantsb}. The figure is analogous to Fig. 8 in Ref. \cite{haxton01} that shows the constraints obtained from different experiments. This figure complements the one that appears in the review of Behr and Gwinner \cite{behr09} as it adds the rubidium numbers to  the constraints obtained from the anapole moment measurement in Cs considering only the experimental uncertainty \cite{wood97,wood99} and the calculations for Fr. 

Figure \ref{constraints} shows the expected constraints for a 3\% measurement in francium and rubidium. The main contribution to the anapole moment in neutron even alkali atoms comes from the valence proton, and the experiment provides a measurement of $g_p$. Using Eq.~\ref{weakconstantsa} we obtain constraints in the direction of the Cs result of Fig. \ref{constraints}a. The fractional experimental uncertainty in $g_p$ is the same as the measurement one. The actual final uncertainty for $g_p$ is considerably higher and depends on the accuracy of the theoretical calculations \cite{flambaum97}. A measurement in any other neutron even alkali atom generates constraints in the same direction as the Cs result, with a confidence band size proportional to the measurement uncertainty. Fig. \ref{constraints}a shows the expected constraint from a 3\% measurement in a neutron even atom (rubidium or francium). The neutron odd alkali atoms have contributions from the valence neutron. The constraints obtained (Eq. \ref{weakconstantsb}) are not colinear to those of Cs and generate a crossing region in Fig. \ref{constraints}. The size of the error band depends on the relative contributions of the valence proton and neutron to the anapole moment which depends on the atom. All the calculations assume $g_p=4$ and $g_n=1$. Fig. \ref{constraints}b shows constraints obtained from measurements in rubidium if the values of $g_n$ and $g_p$ vary. The precise crossing region changes, as well as the constrained area (set in the figure by the error bars). This result shows in a different way the robustness of the proposed measurement. It may be able to give us more than just the DDH coupling constant constraints, but some information of the ratio of $g_n/g_p$.

\begin{figure}
\leavevmode \centering
\includegraphics[width=4in]{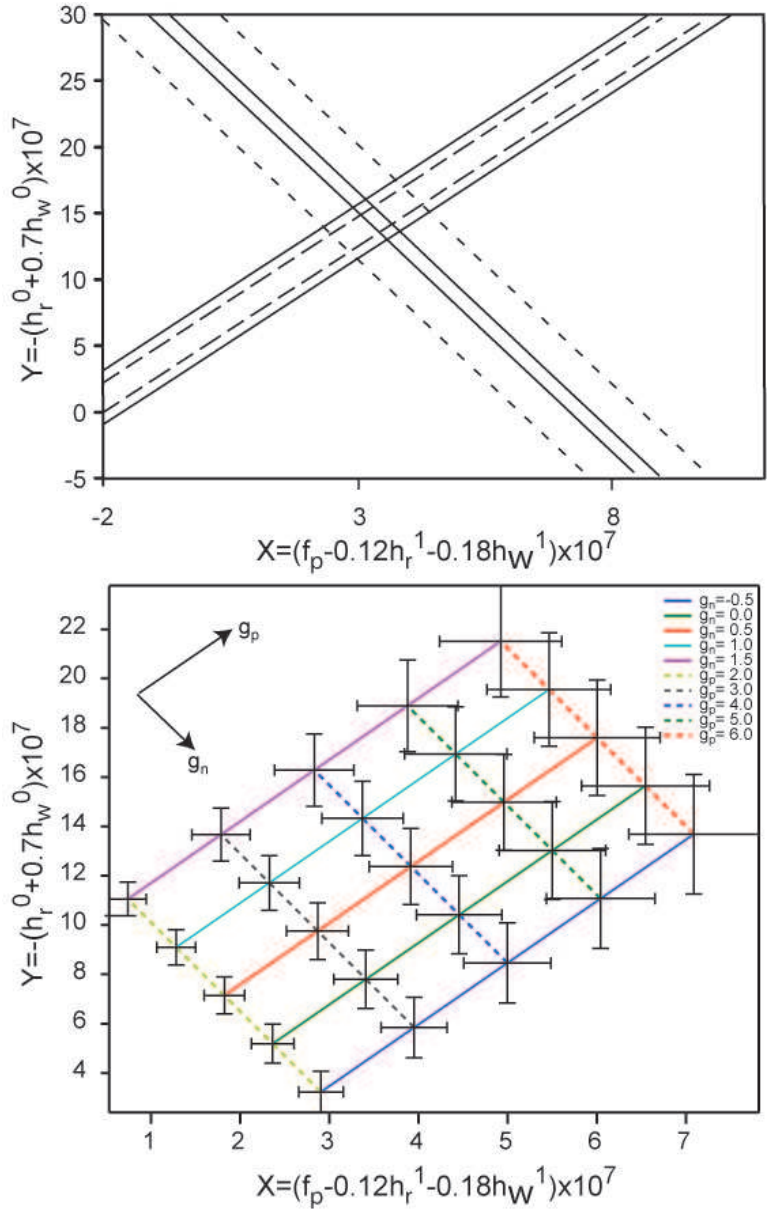}
\caption{(Color online.) a) Constraints to DDH weak meson-nucleon interaction constants from an anapole moment measurement in Rb and Fr. The isotopes with even number of neutrons give constraints aligned with the 14\% Cs result \cite{wood97}, while those with odd number of neutrons give constraints in an different direction. Cs result (small dashed line), Fr 3\% measurement (solid line) and Rb 3\% measurement (long dashed line). b) Sensitivity analysis to the changes in the values of $g_n,~g_p$ by fifty percent. The Error bars mark the 3 \% limits given a set of values. The color lines follow iso-$g_n$ and iso-$g_p$.\label{constraints}}
\end{figure}

\section{Experimental requirements}
This section presents the experimental requirements to measure the anapole moment in chains of Rb and Fr isotopes. Most of the details are in Ref \cite{gomez07}, but here we focus on the differences for Rb and new ways that we have to perform the measurement. The measurement relies on driving an anapole allowed electric dipole ($E1$) transition between hyperfine levels of the ground state. The transition probability is very small and is enhanced by interfering it with an allowed Raman (or magnetic dipole ($M1$)) transition. The excitation is coherent and allows for long interaction times. The signal to noise ratio is linear with time (up to the coherence time). We report on the progress towards an anapole measurement and the possibility of making the measurement in rubidium.

\subsection{Source of atoms}
The work with radioactive atoms requires on-line trapping with an accelerator to have access to reasonably short lifetime isotopes. 
We take the numbers for the production of unstable isotopes from what is available at TRIUMF in the  Isotope Separator and Accelerator (ISAC), where we are an approved experiment. A 500 MeV proton beam collides with an uranium carbide target to produce francium as fission fragments.  A voltage up to 60 kV extracts the atoms as ions from the target. The beam goes through a mass separator and into the trapping area. The yield is up to $2 \times 10^{11}$ s$^{-1}$ for Rb,  and $2 \times 10^{6}$ s$^{-1}$ for Fr, but it is expected to reach $10^8$-$10^9$ for Fr once the accelerator runs at full capacity.

Our current apparatus to go on-line with the accelerator has a vacuum chamber to 
capture atoms in a high efficiency magneto-otpical trap (MOT) operating in batch mode with a neutralizer, as described in Ref.~\cite{aubin03a}. A second science chamber, with controlled electric and magnetic environments for the PNC experiments, connects with the  first through differential pumping. Tests for the transfer of atoms (Rb) between the two show efficiencies above 50\% that allows accumulation of atoms for a longer time.  

\subsection{Measurement}
The anapole moment constant scales with the atomic mass number as $A^{2/3}$ (Eq.~\ref{kappaa}). The anapole induced electric dipole ($E1$) transition amplitude scales faster due to additional enhancement factors \cite{gomez07}. It is about 83 times larger in Fr than in Rb. The magnetic dipole ($M1$) transition amplitude between hyperfine levels, on the other hand, has about the same value for both species. The $M1$ transition gives the main systematic error contribution, and the figure of merit is the ratio of the two transition amplitudes $|A_{E1}/A_{M1}| \sim 1 \times 10^{-9}$ for francium and $|A_{E1}/A_{M1}| \sim 1 \times 10^{-11}$ for rubidium. In order to do a measurement in rubidium it becomes more important to understand and suppress the $M1$ contribution.

The experiment starts by optical pumping all the atoms to a particular level $|F_1,m_1 \rangle$. The atoms interact with two transitions for a fixed time. One corresponds to a Raman transition (parity conserving) and the other to the anapole induced electric dipole ($E1$) transition (parity violating). Both are resonant with the ground state hyperfine transition $|F_1,m_1 \rangle \to |F_2,m_2 \rangle$ in the presence of a static magnetic field. This triad of vectors defines the coordinate axis. The interference of both transitions gives a signal linear in the anapole moment \cite{gomez07}. The field driving the $E1$ transition is inside a microwave Fabry-Perot cavity. The signal to measure is the population in $|F_2,m_2 \rangle$ at the end of the interaction with a given handedness of the system.

The specific magnetic field and transition levels reduce the sensitivity to fluctuations. There is an operating point where the transition frequency varies quadratically with the magnetic field for a $|\Delta m|=1$ transition. The operating field grows with the hyperfine separation and it is larger in Fr than in Rb. Table \ref{tableparameters} shows the magnetic field for different isotopes of Fr and Rb. The transition $m_1=0.5 \to m_2=-0.5$ in odd neutron isotopes has a small magnetic field. The electronic contribution to the linear Zeeman effect cancels for the two levels at low magnetic fields, but the nuclear magnetic contribution remains since the two states belong to different hyperfine levels. The magnetic field for odd neutron Fr isotopes is smaller than previosly reported ($\sim$ 2000 G) \cite{gomez07}. 

\begin{table}
\leavevmode \centering \caption{Operating point for the magnetic field ($B_0$), resonant frequency ($\nu_m$) and Zeeman sublevels ($m_1$, $m_2$) for the transition.}
\begin{tabular} {cccccccc}
~~ Atom ~~ & ~~ Isotope ~~ & ~~ Spin ~~ & ~~ $m_1$ ~~ & ~~ $m_2$ ~~ & ~~ $B_0$ (G) ~~ & ~~ $\nu_m$ (Mhz) ~~ \\
\hline
Rb	&	84		& 2		& -0.5	& 0.5		& 0.2			& 3084        \\
		&	85		& 5/2	& 0			& -1		& 186.1		& 2992        \\
		&	86		& 2		& 0.5		& -0.5	& 0.3			& 3947        \\
		&	87		& 3/2	& 0			& -1		& 654.2		& 6602        \\
		&	88		& 2		& 0.5		& -0.5	& 0.03		& 1191        \\
\hline
Fr	&	208		& 7		& 0.5		& -0.5	& 3.3			& 49880        \\
		&	210		& 6 	& 0.5		& -0.5	& 3.4			& 46768        \\
		&	212		& 5		& 0.5		& -0.5	& 4.5			& 49853        \\
\end{tabular}
\label{tableparameters}
\end{table}

The dimensions of the microwave cavity scale with the wavelength of the transition ($\lambda_m \sim 6$ mm for Fr and $\lambda_m \sim 6$ cm for Rb). The mirror separation of the Fabry-Perot cavity should be at least 20 cm for Rb. The anapole signal remains unchanged between Fr and Rb by putting more power in the microwave cavity to compensate for the loss in the nuclear size.

We hold the atoms in place for the measurement using a far off resonance trap (FORT) \cite{friedman02}. Changes in the FORT wavelength allow its use for both rubidium and francium. The dipole trap causes an ac Stark shift that is different for the two hyperfine levels. The differential shift causes a change of the resonant frequency and eventually leads to decoherence. The Stark shift in a FORT is in the same direction for both hyperfine levels but of different size due to the different detuning. The differential shift is proportional to the hyperfine splitting, and it is reduced by an order of magnitude in rubidum.

The measurement in both species depends on the effectiveness of the suppression mechanisms \cite{gomez07}. The first suppression mechanism works by having the atoms in the magnetic field node (electric field antinode). The reduction depends on the magnitude of the field at the edges of the atomic cloud. Since the wavelength increases by an order of magnitude in rubidium, the suppression improves by the same amount. The second suppression mechanism works by forcing the $M1$ transition to have the wrong polarization for the levels considered. It remains unchanged in rubidium. The atoms oscillate around the magnetic field node for the third suppression mechanism. The suppression is proportional to the $M1$ field, and since it gets reduced because of the better positioning in the node, we can gain an order of magnitude (assuming no increase in the driving field power). The suppression mechanisms work better in rubidium than in francium by two orders of magnitude because of better positioning to the magnetic field node. The improvement compensates the two orders of magnitude loss in the figure of merit ($|A_{E1}/A_{M1}|$). 

We compare the requirements in rubidium to those established on Table III of Ref. \cite{gomez07} for francium. We assume an increase in the microwave power to keep the same $E1$ transition amplitude. The magnetic field stability is still about $10^{-5}$ but since now the magnetic field is smaller this means that the field has to be controlled to about 10 $\mu$G. The requirements on all the systematic effects that depend on the $M1$ transition produced by the microwave cavity increase by two orders of magnitude. This is because by increasing the microwave field we increase the $E1$ and $M1$ transition amplitude by the same amount. The systematic effects introduced by the dipole trap or Raman beams remain the same.

\section{Optical Dipole trap}

We report on the experimental implementation of the optical dipole trap in the science chamber. The dipole trap design aims to decrease the photon scattering and differential ac Stark shift introduced by the laser forming the trap \cite{chin01,romalis99}.
We use a FORT to reduce the photon scattering rate. The ac Stark shift depends on the position in the trap and the atomic state. The shift changes with time as the atoms move in the trap. We choose a blue detuned trap  where the atoms are confined on the dark region of the trap.

We use a rotating dipole trap because we can control the shape and size dynamically. A laser rotating faster than the motion of the atoms creates a time averaged potential equivalent to a hollow beam potential \cite{friedman00}. The laser beam propagating in the $z$ direction goes through two acousto-optical modulators (AOMs) placed back-to-back in the $x$ and $y$ directions respectively. We use the beam that corresponds to the first diffraction order in both directions, the (1,1) mode. We scan the modulation frequency of both AOMs with two phase-locked function generators (Stanford Research SR345) to generate different hollow beam shapes.

The general expression of the time averaged potential $U({\rho,z})$ in the radial direction for linearly polarized light and a detuning larger than the hyperfine structure splitting, but smaller than the fine structure splitting  is \cite{kuppens00}:
\begin{equation}
U({{\rho,z}})=\frac{\hbar \gamma}{24 I_S}  \left[ \frac{1}{\delta_{1/2}}+\frac{2}{\delta_{3/2}} \right] 
\frac{{\oint}_{{\rho}'\in\textbf{l}}I(\rho-{\rho}',z)dl}{{\oint}dl}
\end{equation}
where $\gamma$ is the natural linewidth, $I_S$ is the saturation intensity defined as $I_S=2\pi^2 \hbar c \gamma/(3\lambda^3)$, and $I(\rho,z)$ is the Gaussian beam intensity at position $(\rho,z)$. The integral over the contour of the rotating laser beam ${\bf l}$ gives the time averaged potential. The detunings $\delta_{1/2}$ and $\delta_{3/2}$ in units of $\gamma$.

Tightly focusing the laser at the position of the atoms confines them along the beam axis. Fig. \ref{prog:ax} shows the shape of the potential both along the radial and axial directions for a circular shaped trap.
\begin{figure}[hpt]
\centering
\includegraphics[width=5in]{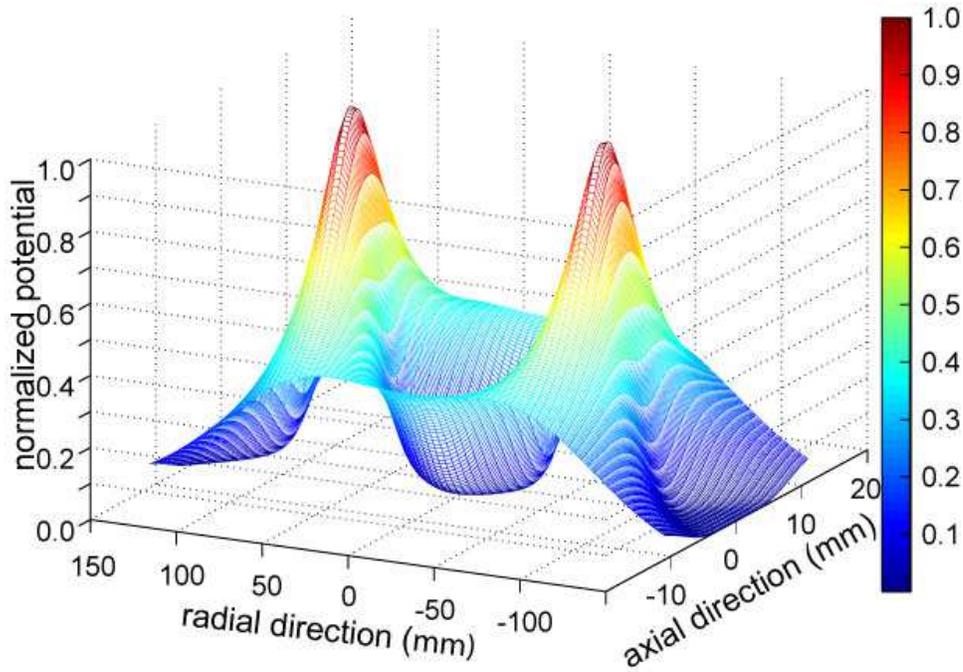}
\caption{\label{prog:ax} (Color online) Time averaged potential along the axial and radial directions for a cigar shaped trap. The aspect ratio is not to scale in the figure.}
\end{figure}

We study the lifetime and spin relaxation rate of such a circular trap with a beam of 400 mW and blue detuned 2.5 nm from the $^{87}$Rb $D_2$ line. The spin relaxation is critical for the anapole measurement. The beam rotating frequency is 50 kHz, which is much faster than the oscillation frequency of the trap ($\leq 1$ kHz). The trap has a transverse diameter of 150 $\mu$m, an axial diameter of 24 mm, and a potential depth of 60 $\mu K$ (normalized potential value of 0.4 in Fig~\ref{prog:ax}). We measure the atom number in the dipole trap after a pre-set hold time by shinning a 100 $\mu$s long resonant pulse and imaging the fluorescence into a photomultiplier tube (Fig. \ref{prog:life}a). We image the fluorescence from a region of radius of 2 mm. We see a rapid decay during the first 100 ms from fast atoms that can not be confined in the radial direction. The rapid decay is followed by a slower decay (2.5 s lifetime). The slow decay is shorter than the MOT lifetime (30 s) and corresponds to the continous diffusion of the atoms out of the imaging region. This is supported by the calculation shown in Fig.~\ref{prog:life}a that gives the remaining number of atoms in the imaging area using the expected temperature of the atoms. We follow the method of Ref. \cite{ozeri99} to measure the spin relaxation rate. We load the atoms from a magneto-optical trap (MOT) to the dipole trap, turn off the magnetic field and MOT beams and pump the atoms to the $F=1$ ground state. We get the relaxation rate due to the interaction of the atoms with the dipole trap laser by comparing the populations of the atoms in both hyperfine levels as a function of time. Figure \ref{prog:life}b shows the fraction of atoms in the $F=2$ ground state as a function of time. An exponential fit to the data gives a spin relaxation time of 840 ms, similar to previous measurements \cite{ozeri99}. This is a first step towards the 20 ms coherent interaction of the proposed data taking cycle in Ref. \cite{gomez07}
\begin{figure}[hpt]
\centering
\includegraphics[width=4in]{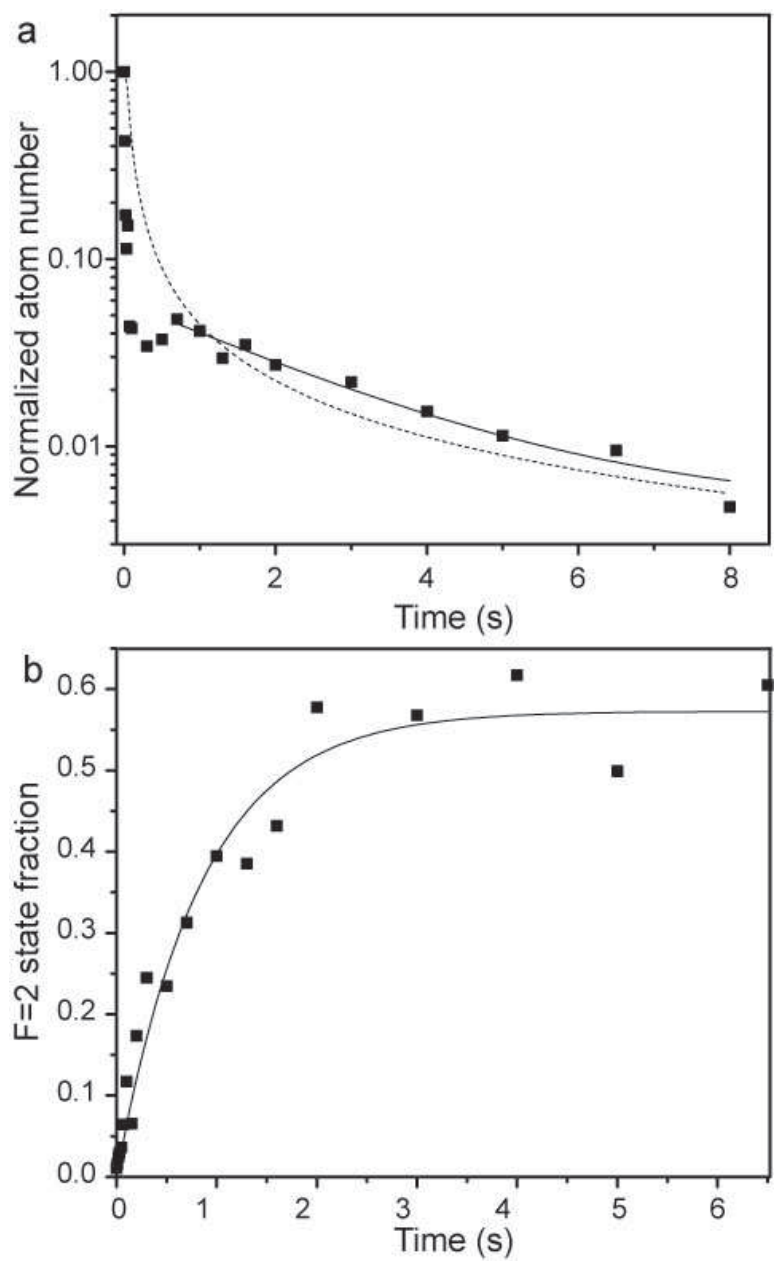}
\caption{\label{prog:life}a) Lifetime measurement of the atoms in the dipole trap, filled squares experimental data, dashed line atoms escaping model, continuous line exponential fit. b) Measurement of the spin relaxation time by plotting the fraction of the atoms in the $F=2$ state in the dipole trap. The continuous line is the exponential fit (lifetime 840 ms).}
\end{figure}

We reduce the diffusion of the atoms in the axial direction by adding a one-dimensional (1D) blue denuned standing wave with a frequency different from the one used for the rotating trap. The combination of tight radial confinement from the rotating trap and confinement in the axial direction from the standing wave gives a higher density dipole trap (Fig. \ref{prog:bt}). It also opens the possibility to study the motion of atoms in 2D billiards with arbitrary transverse shape \cite{friedman01}. 

\begin{figure}[hpt]
\centering
\includegraphics[width=4in]{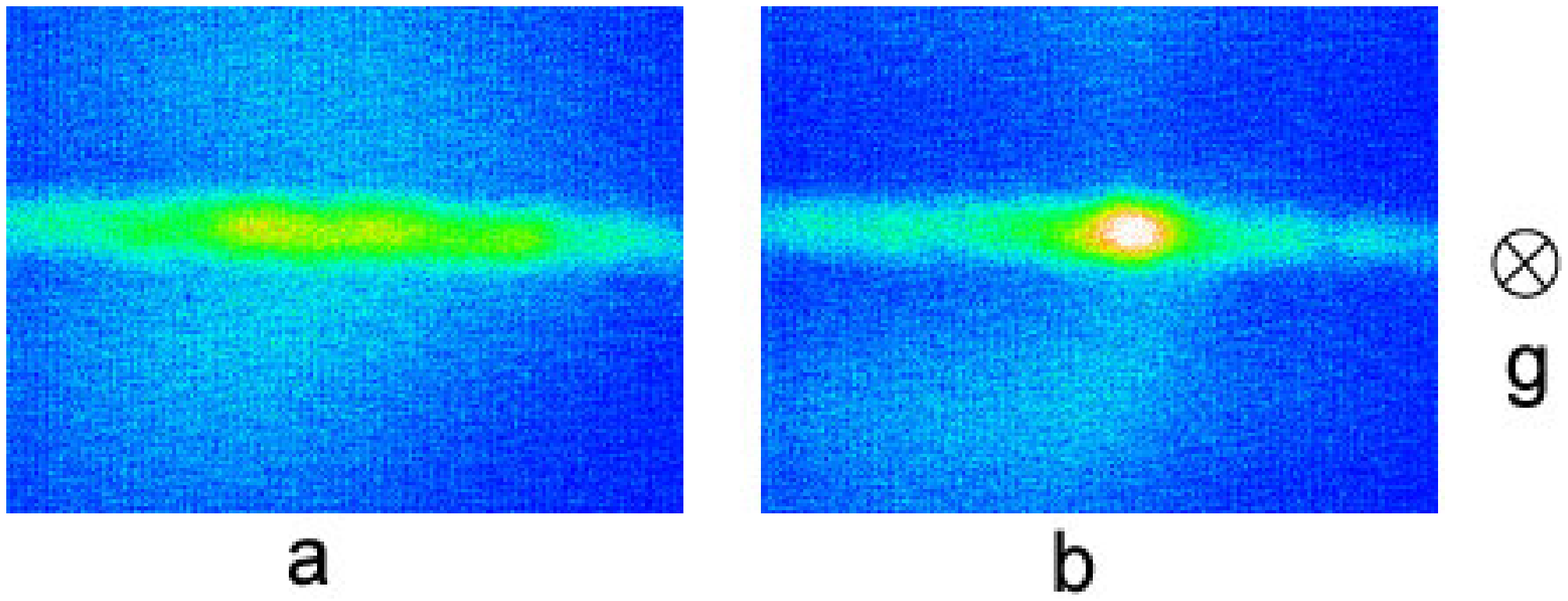}
\caption{\label{prog:bt} Fluorescence image of the atoms 35 ms after turning off the magnetic field and MOT beams. a)
rotating dipole trap. b)  rotating dipole trap and 1D blue detuned standing wave. Gravity (g) goes into the paper in the figure.}
\end{figure}

The symmetry of the trap is an important point for the anapole moment measurement. As we scan the beam, there will be diffraction power changes on the AOMs. This has been pointed out in the study with Bose-Einstein condensates where the uniformity is required to avoid parametric excitation \cite{williams08}. We feedforward on the RF power to reduce the diffraction variations \cite{schnelle08}. Fig. \ref{prog:pl} shows the increased stability in the diffraction power as we rotate the beams using this method.
\begin{figure}[hpt]
\centering
\includegraphics[width=4in]{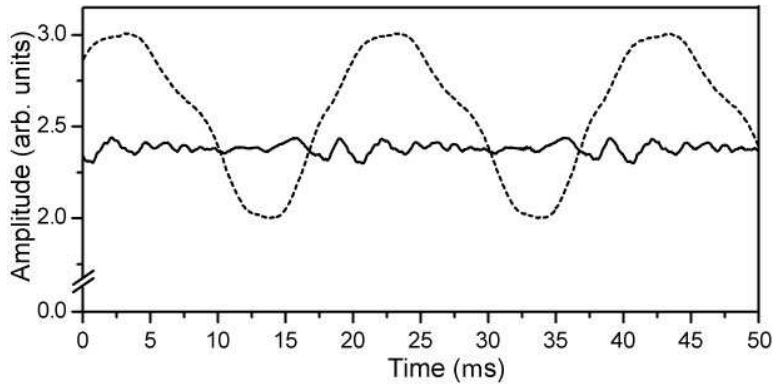}
\caption{\label{prog:pl} Intensity profile of the diffracted light showing a few oscilation periods that generate the trap with (solid line) and without (dashed line) feedforward on the RF power to the AOMs.}
\end{figure}

\section{Conclusions}

The measurement of the anapole moment in a chain of isotopes can constraint the values of the DDH weak meson-nucleon interaction constants. The measurement of the anapole moment is possible in any of the heavy alkali atoms (rubidium, cesium or francium) but it becomes increasingly difficult with decreasing atomic number. The anapole moment for the two Rb naturally available isotopes has the opposite sign which can be useful for the study of systematic effects. Neutron odd isotopes have transitions insensitive to magnetic field fluctuations at small static values of the magnetic field. We have a working  rotating blue detuned dipole trap necessary to hold the atoms for the duration of the anapole moment measurement. The trap shows a spin relaxation time of 840 ms. The dipole trap will be used in future measurements in both francium and rubidium.

\section*{Acknowledgments}

Work supported by NSF. E. G. acknowledges support from CONACYT (cooperacion bilateral). We would like to thank J. A.  Behr, B. A. Brown, N. Davidson,  S. Schnelle, G. D. Sprouse, for helpful discussions and A. Perez Galvan for his work on the apparatus and interest on this project.

\end{document}